\documentclass[pre,twocolumn,floatfix,amsmath,amsfonts]{revtex4}

\def\({\left(}
\def\){\right)}

\usepackage{epsfig}
\usepackage{graphicx}
\usepackage{amsmath}
\usepackage{amssymb}
\usepackage{dcolumn}
\usepackage{bm} 
\usepackage{dsfont, srcltx}
\usepackage{color}

\begin{document}

\title{Discrete breathers assist energy transfer to ac driven nonlinear chains}

\author{D.~Saadatmand$^{1}$}
\email{saadatmand.d@gmail.com}

\author{Daxing Xiong$^{2}$}
\email{phyxiongdx@fzu.edu.cn}

\author{V.~A.~Kuzkin$^{3,4}$}
\email{kuzkinva@gmail.com}

\author{A.~M.~Krivtsov$^{3,4}$}
\email{akrivtsov@bk.ru}

\author{A.~V.~Savin$^{5}$}
\email{dmitriev.sergey.v@gmail.com}

\author{S.~V.~Dmitriev$^{6,7}$}
\email{dmitriev.sergey.v@gmail.com}

\affiliation{$^1$Department of Physics, University of Sistan and
Baluchestan, Zahedan, Iran
\\
$^2$Department of Physics, Fuzhou University, Fuzhou 350108,
Fujian, China
\\
$^3$Peter the Great Saint Petersburg Polytechnical University, Polytechnicheskaya st. 29, Saint Petersburg, Russia
\\
$^4$
Institute for Problems in Mechanical Engineering RAS, Bolshoy pr. V.O. 61, Saint
Petersburg, Russia
\\
$^5$
Semenov Institute of Chemical Physics, Russian Academy of Science, Moscow, 119991, Russia
\\
$^6$Institute for Metals Superplasticity Problems RAS, Khalturin
39, 450001 Ufa, Russia
\\
$^7$National Research Tomsk State University,
Lenin Ave, 36, 634050 Tomsk, Russia
 }

\begin{abstract}
One-dimensional chain of pointwise particles harmonically coupled with nearest neighbors and placed in six-order polynomial on-site potentials is considered. Power of the energy source in the form of single ac driven particles is calculated numerically for different amplitudes $A$ and frequencies $\omega$ within the linear phonon band. The results for the on-site potentials with hard and soft nonlinearity types are compared. For the hard-type nonlinearity, it is shown that when the driving frequency is close to (far from) the {\em upper} edge of the phonon band, the power of the energy source normalized to $A^2$ increases (decreases) with increasing $A$. In contrast, for the soft-type nonlinearity, the normalized power of the energy source increases (decreases) with increasing $A$ when the driving frequency is close to (far from) the {\em lower} edge of the phonon band. Our further demonstrations indicate that, in the case of hard (soft) anharmonicity, the chain can support movable discrete breathers (DBs) with frequencies above (below) the phonon band. It is the energy source quasi-periodically emitting moving DBs in the regime with driving frequency close to the DBs frequency, that induces the increase of the power. Therefore, our results here support the mechanism that the moving DBs can assist energy transfer from the ac driven particle to the chain.
\end{abstract}
 \maketitle

\section {Introduction}

For many physical systems, a common basic problem is the response of a nonlinear medium to periodic excitations at the boundary or inside a local region in the bulk~\cite{Rossler,GL2002,Rings2004,JSWang,LZHL,Johansson,AiHeHu,Beraha,Junior,Evazzade}. The energy can flow or not flow from the energy source into the medium, depending on the medium, the frequency, and the amplitude of the excitations. Linear medium absorbs energy only if the frequency of the source is within the spectrum of small-amplitude running waves (phonons) supported by the medium. While for a nonlinear medium, the energy source can transmit energy into the medium even at driving frequencies outside the small-amplitude phonons spectrum. According to the so-called supratransmission effect~\cite{GL2002,Aranson,Caputo,Khomeriki}, energy, in this case, is transported by the moving discrete breathers (DBs) \cite{Dolgov,ST1988,FW1998,FW2008,Uspekhi} when the driving amplitude is above a threshold value. But some new phenomena beyond supratransmission effect can also been observed, e.g., for excitation frequencies outside the phonon spectrum, energy can flow into a nonlinear discrete system even at small driving amplitudes~\cite{Evazzade}; when the system is in contact with heat baths~\cite{AiHeHu}, the amplitude threshold for this nonlinear supratransmission effect will be absent.

Recently, interest to the energy transport by linear and nonlinear phonons and by DBs has increased enormously, due to the emerging new field of {\em phononics}~\cite{Phononics,Phononics1} and the recent theoretical and experimental progress on anomalous heat transport in low-dimensional systems~\cite{Lepri,Dhar,Anomaly1,Anomaly2,Anomaly3,Anomaly4,Anomaly5,Anomaly6,Anomaly7,Anomaly8}, on thermal diodes~\cite{Diod1,Diod3,Isotops,Diod2,Diod4}, on thermal transistors~\cite{Transist1,Transist2}, and on various thermal logic gates~\cite{Gates1,Gates2,Gates3}. On the one hand, the relevant theoretical studies showed that DBs~\cite{XZ2016,XSD2017} and solitons~\cite{JYZZ2017} can affect thermal conductivity in nonlinear chains. Randomly distributed defects~\cite{SKH2010, LeKr2007,Isotops} can also influence heat transport. In particular, the study of~\cite{SavKos} revealed that heat transport is normal (obeying the Fourier law) in the chains with the interatomic potentials allowing breaking of interatomic bonds. The work~\cite{Gradient} suggested the properties of phonon localization and thermal rectification in the chains with strain gradient.

On the other hand, external periodic driving gives new ways in manipulating energy flux in nonlinear lattices. Heat can flow from the low-temperature to the high temperature heat bath in nonlinear lattices when the temperature of a heat bath is time-periodically modulated~\cite{LZHL} or when a driving force with frequency in a certain range is applied at the lattice boundary~\cite{AiHeHu}. Experimental setup for low-frequency phonon cooling with external periodic driving in diamond nanoresonator has been proposed in~\cite{Kepesidis1,Kepesidis2}.

In fact, more efforts has been devoted to relating the above two aspects. With the time-dependent frequency driving of a small amplitude traveling wave, excitation of solitons in the Korteweg-de Vries equation has been analyzed~\cite{Aranson}. The possibility of optical excitation of DBs in crystals has also been demonstrated~\cite{Rossler}. Excitation of standing DBs with time-modulated vibration amplitude in a strained graphene has been observed~\cite{Evazzade}.  

All of the above studies indicate that it is very interesting to study the basic mechanism for energy transfer in nonlinear lattices by external ac driving, and to demonstrate the relevant roles of nonlinear excitations, such as solitons and DBs, within this process. However, so far, the excitation of the latter case, i.e., the DBs, has only been investigated for driving frequencies {\em outside} the phonon band~\cite{GL2002,Rings2004,Evazzade,Aranson,Caputo,Khomeriki}. In this work we therefore study energy transfer in 1D nonlinear chains in the case when driving frequency is {\em within} the phonon band. 

The rest of this work is composed as follows. In Sec.~\ref{Sec:IIA} the focused 1D nonlinear chains are described, the linear phonon spectrum of these chains is discussed, and the details of our investigation scheme are briefly presented. In Sec.~\ref{Sec:III} the energy transfer with harmonically driven particle in the middle of a linear chain is first analytically demonstrated. Then, numerical results for the chains with hard- and soft-type nonlinearities are given. In Sec.~\ref{Sec:IV} the properties of standing and moving DBs in the cases of hard- and soft-type nonlinearity are discussed to explain the Sec.~\ref{Sec:III}'s findings. Finally, Sec.~\ref{Sec:V} draws our conclusion.

\section {Models} \label{Sec:IIA}

\begin{figure}
\includegraphics[width=8.0cm]{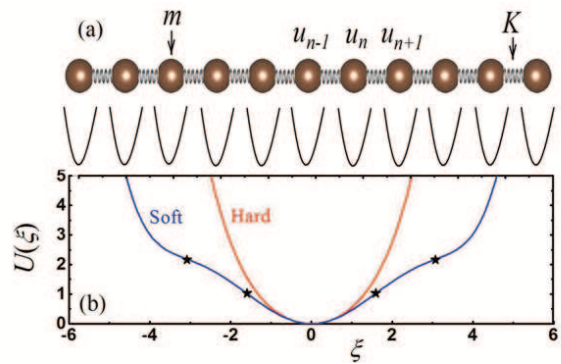}
\caption{(a) Schematic plot of the 1D chain of harmonically coupled point-wise particles in the anharmonic on-site potential. (b) The on-site potential as a function of $\xi$ for hard-type (red line) and soft-type (blue line) potentials. The
stars show the inflection points in the soft-type case. }
\label{fig1}
\end{figure}

We consider the 1D chains [see Fig.~\ref{fig1}(a)] of pointwise (but with mass $m$) particles whose Hamiltonian is defined by 
\begin{equation}\label{Hamiltonian}
H=\sum_n\left[\frac{m\dot{u}_n^2}{2}+V(u_{n+1}-u_{n})+U(u_{n})\right],
\end{equation}
where $u_n$ is the displacement of the $n$th particle from its equilibrium position, $\dot{u}_n$ is its velocity (overdot means derivative with respect to time), and
\begin{equation}\label{PotentialLinear}
V(\xi)=\frac{K\xi^2}{2}
\end{equation}
is the harmonic potential with stiffness constant $K$ describing the interaction of each particle with its nearest neighbors. For the on-site potential we take
\begin{equation}\label{PotentialHard}
U(\xi)=k \xi^2+\alpha \xi^4+\beta \xi^6,
\end{equation}
where $k$ is the coefficient in front of the harmonic term, while $\alpha$ and $\beta$ are the coefficients that define the contributions from the quartic and six-order terms, respectively.

Without the loss in generality we set $m$=1, $K=1$. Our attention then is focused on the on-site potential, where we take $k=1/2$ and $\beta=1/720$. For $\alpha$, we consider the following two cases, i.e., $\alpha=1/24$ for the hard-type anharmonicity and $\alpha=-1/24$ for the soft-type (only for not very large $\xi$ will be taken into account). Thus, in both cases we have unbounded on-site potential, as demonstrated in Fig.~\ref{fig1}(b). In fact, for very large $\xi$, both potentials are of hard-type since the leading term is proportional to $\xi^6$. As shown in Fig.~\ref{fig1}(b), the soft-type potential has four inflection points at $\xi\approx\pm 1.59$ and $\xi\approx\pm 3.07$, which are shown by stars. Therefore, in this study we do not consider excitations with the displacements of particles exceeding the first inflection point to ensure that the anharmonicity of the on-site potential is really of soft type. For displacements exceeding the second inflection point the on-site potential will become effectively harder due to the effect of the six-order nonlinear term.

From Eqs.~\eqref{Hamiltonian}-\eqref{PotentialHard} the following equations of motion can be
derived
\begin{equation}\label{EMo}
m\ddot{u}_n=
K(u_{n-1}-2u_n+u_{n+1}) - 2ku_n -4\alpha u_n^3 - 6\beta u_n^5.
\end{equation}
As to the case of small amplitude vibrations, the forth- and sixth-order nonlinear terms can be neglected, and thus 
\begin{equation} \label{EquationMotion}
m\ddot{u}_n=K(u_{n-1}-2u_{n}+u_{n+1})-2ku_{n}.
\end{equation}
The solutions of the above equation are the linear combinations of
normal modes $u_n \sim \exp [i (q n -\omega_q t)]$ with
wave number $q$ and frequency $\omega_q$ following the dispersion relation:
\begin{equation}\label{Dispersion}
\omega_q^2=\frac{2}{m}[k-K(\cos q-1)].
\end{equation}

\begin{figure}
\includegraphics[width=8.0cm]{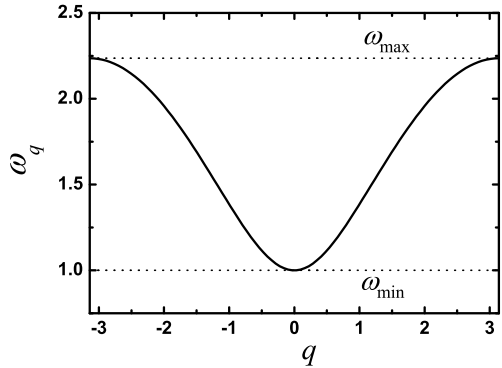}
\caption{Dispersion relation for small-amplitude waves (phonons) supported by the considered chain of particles. The phonon band ranges from $\omega_{\min}=1$ to $\omega_{\max}=\sqrt{5}\approx 2.236$.}
\label{fig2a}
\end{figure}

In Fig.~\ref{fig2a} the dispersion relation~\eqref{Dispersion} is shown within the first Brillouin zone. It suggests that the systems support the small-amplitude running waves (phonons) with frequencies ranging from $\omega_{\min}=1$ to $\omega_{\max}=\sqrt{5}\approx 2.236$. 
Phonon's group velocity defined by $v_g=\rm{d} \omega_{\mit{q}}/\rm{d} \mit{q}$ vanishes for $q\rightarrow 0$ and $q\rightarrow \pm\pi$. While the phonons with fastest velocity is the one having frequencies in the middle of the phonon band. This fact will be used in the discussion of the power of the energy source.


In simulations we usually consider chains of particles $N=4000$. This size is long enough to avoid the effect of boundaries on energy transfer from the driven particle within the simulation run of $t_{\max}=2000$ time units. The periodic boundary condition is used, i.e., to set $u_n=u_{n+N}$. Equations of motion are integrated with the help of the St\"ormer method of order six with the time lag $\tau=0.005$. Further decrease of the time step did not affect the final simulation results.

\section {Energy transfer to ac driven chain} \label{Sec:III}

Initially, in the middle of the chain, one particle is forced to move according to the harmonic law
\begin{equation}\label{DrivingForce}
u_{N/2}=A\sin(\omega t), \quad T=\frac{2\pi}{\omega},
\end{equation}
with the driving parameters, amplitude $A$, frequency $\omega$, and period $T$. All other particles are set with zero displacements and velocities. The driven particle can be regarded as the energy source. The chain can accept or not accept the energy from the source depending on $A$ and $\omega$. 

The driving~\eqref{DrivingForce} is applied during the whole simulation run up to $t_{\max}=2000$. Such choice of $t_{\max}$ ensures that the perturbation from the energy source does not reach the boundaries of the chain for any driving parameters. At the end of the simulation run, we then calculate the total (kinetic+potential) energy $E$ of the chain, and finally find the averaged power of the energy source over all simulation runs as follows
\begin{equation}\label{Power}
P=\frac{E(t_{\max})}{t_{\max}}.
\end{equation}

Following this way, while, the power of the energy source is actually a function of time. To obtain a detailed characteristic of the source power, in practice, we choose to calculate the total energy of the chain $E_j$ at times $t_j=jT$ with $j=0,1,2,...$, and $T$ the driving period. With this information, we then calculate the power averaged over each driving period by

\begin{equation}\label{PowerT}
p_j=\frac{E_{j+1}-E_j}{T}.
\end{equation}

\subsection {Exact solution for driven harmonic chain} \label{Sec:IIIA}
Before starting to discuss the nonlinear cases, usually, it is very instructive to analyze the behavior of a linear system first~\cite{Linear1,Linear2,Linear3,Linear4,Linear5, Linear6}. In this subsection, we  derive an exact expression for the total energy, $E$, of the linear system~\eqref{EquationMotion} subjected to external excitation~\eqref{DrivingForce} under zero initial conditions. This allows us to find both the power $P$ defined by Eq.~\eqref{Power} and the power for each driving period defined in Eq.~\eqref{PowerT}.

To do this we introduce a new variable
\begin{equation}
  w_n = u_n - A \sin(\omega t)
\end{equation}
for convenience. This new variable $w_n$ then satisfies the following equations
\begin{equation}\label{eq w_n}
\begin{array}{l}
\displaystyle
m\ddot{w}_n = K(w_{n+1} - 2 w_n + w_{n-1}) -2k w_n + \\ [4mm]
+A(m\omega^2 - 2k)\sin(\omega t), \quad w_{\frac{N}{2}} = w_{\frac{3N}{2}}=0
\end{array}
\end{equation}
with initial conditions
\begin{equation}\label{IC w_n}
 w_n = 0, \qquad \dot{w}_n = -A\omega, \qquad n = \frac{N}{2}+1,..,\frac{3N}{2}-1.
\end{equation}

Normal modes for Eqs.~(\ref{eq w_n}) are~$\sin\frac{\pi j(2n-N)}{2N}$. Corresponding eigen frequencies are  calculated as
\begin{equation}{}
\Omega_j^2 = \omega_{\rm min}^2 + \frac{2K}{m} \(1- \cos\frac{\pi j}{N}\),  \qquad \omega_{\rm min}^2 =\frac{2k}{m}.
\end{equation}
Therefore, the exact solution of Eqs.~(\ref{eq w_n}) can be represented as a linear combination of all normal modes. Including the initial conditions yields
\begin{equation}\label{sol w_n}
 \begin{array}{l}
 \displaystyle w_n =  \frac{A}{N}\sum_{j=1}^{N-1}
 B_j
 \left[(\omega^2 - \omega_{\rm min}^2)\sin(\omega t) - \right.
 \\[4mm]
 \displaystyle \left. \frac{\omega}{\Omega_j}(\Omega_j^2-\omega_{\rm min}^2) \sin(\Omega_j t)\right]
 \sin \frac{\pi j(2n-N)}{2N},
 \\[4mm]
 \displaystyle B_j = \frac{(1-(-1)^j){\rm ctg}\frac{\pi j}{2N}}{(\Omega_j^2-\omega^2)}.
 \end{array}
\end{equation}
%
To summarize, formula~(\ref{sol w_n})  is the exact solution of Eqs.~(\ref{eq w_n}) under initial conditions~(\ref{IC w_n}).

Now let us turn to the total energy of the linear system, which is calculated by using the law of energy balance:
\begin{equation}\label{eq E}
 E(t) = A \omega \int_0^t f(\tau) \cos(\omega \tau) {\rm d}\tau,
\end{equation}
where $f$ is the force driving the particle number~$N/2$. According to the second Newton's law, force~$f$ is equal to the difference between acceleration of this particle and forces induced by the neighboring particles and the on-site potential. Then  
\begin{equation}\label{eq f}
 f(t) = A(2k  + 2K - m\omega^2) \sin(\omega t) -2K u_{\frac{N}{2}+1}.
\end{equation}
Here the identity $u_{\frac{N}{2}+1} = u_{\frac{N}{2}-1}$ is used. This identity follows from symmetry of the problem with respect to particle number~$N/2$.  

Next, substituting Eqs.~(\ref{sol w_n}), (\ref{eq f}) into Eq.~(\ref{eq E}) and performing integration, yields
\begin{equation}\label{sol E}
\begin{array}{l}
\displaystyle \frac{E(t)}{A^2}= \frac{1}{2}(2k  -m\omega^2) \sin^2(\omega t) -\frac{K}{N}
 \sum_{j=1}^{N-1} B_j  g_j \sin \frac{\pi j}{N},
 \\[4mm]
\displaystyle g_j  =  (\omega^2 - \omega_{\rm min}^2)\sin^2(\omega t)
 -
 \frac{2\omega^2(\Omega_j^2-\omega_{\rm min}^2)}{\Omega_j(\Omega_j^2-\omega^2)} h_j,
\\[4mm]
\displaystyle h_j = \Omega_j -\omega \sin(\Omega_j t)\sin(\omega t) - \Omega_j \cos(\Omega_j t)\cos(\omega t),
\end{array}
\end{equation}
which is an exact expression for the energy of the linear chain at any moment in time.

Using Eq.~(\ref{sol E}), one then can calculate both the averaged power of the energy source over time from $t=0$ to $t=t_{\max}$ and the power for each driving period, based on Eq.~(\ref{Power}) and Eq.~(\ref{PowerT}), respectively. For large $N$, sums in formula~\eqref{sol E} can be replaced by integrals. Then applying asymptotic methods, we show that at large times the expression for $E$ has simple form
\begin{equation}\label{E appr}
\frac{E(t)}{A^2} \approx \frac{1}{2} m \omega^2 c_g(\omega) t,
\end{equation}
where~$c_g = \frac{1}{\omega}\sqrt{\(\omega^2-\omega_{\rm min}^2\) \(\omega_{\rm max}^2 - \omega^2\)}$ is the group velocity. 

Formula~\eqref{E appr} has very transparent physical meaning, since $mA^2\omega^2/2$ is the energy density of phonons with amplitude $A$ and frequency $\omega$, while $c_gt$ is the distance traveled by the phonons at time $t$. Thus, the power of the energy source at large times is just a product of phonons energy density and phonons group velocity.

\subsection {Numerical results for driven nonlinear chain} \label{Sec:IIIB}

\begin{figure}
\includegraphics[width=9.5cm]{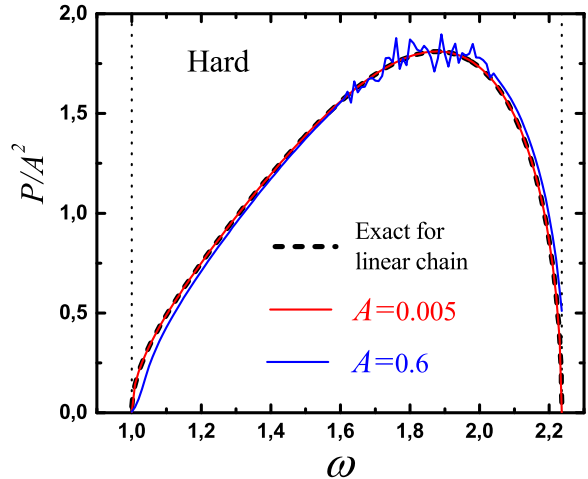}
\caption{Power $P$ of the energy source, normalized by $A^2$, as a function of driving frequency for two driving amplitudes, $A=0.005$ and 0.6. The exact result for the linear chain is shown by the thick dashed line. Vertical dashed lines show the edges of the phonon band
with $\omega_{\min}=1$ and  $\omega_{\max}=\sqrt{5}$.} \label{fig2}
\end{figure}

\begin{figure}
\includegraphics[width=10cm]{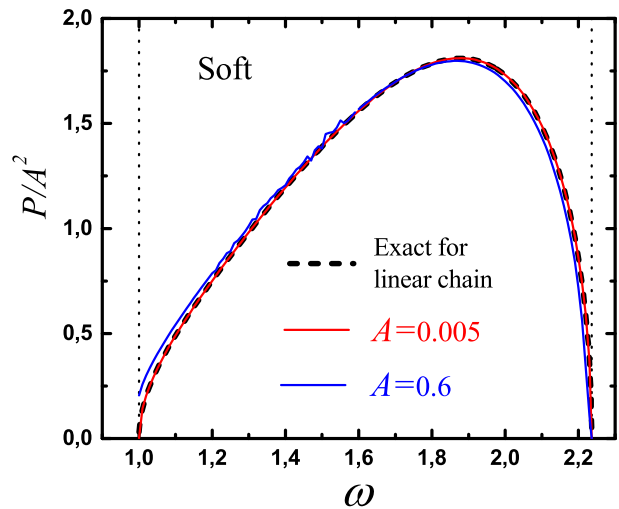}
\caption{The same result as that shown in Fig.~\ref{fig2} but for the case of soft
nonlinearity.} \label{fig3}
\end{figure}

In Figs.~\ref{fig2} and~\ref{fig3} the power $P$ of the energy source, normalized by $A^2$, is plotted as a function of the driving frequency for two driving amplitudes, $A=0.05$ and $A=0.6$, for the hard-type ($\alpha=1/24$) and the soft-type ($\alpha=-1/24$) nonlinearities, respectively. Driving frequencies are set within the phonons band and marked by the vertical dashed lines.  Here, remind that, as mentioned, $P$ is the averaged power over the whole numerical run of $t_{\max}=2000$. For comparison, the derived exact solution [one can substitute Eq.~(\ref{sol E}) into Eq.~(\ref{Power})] and some numerical results for several small driving amplitudes are plotted together. For small $A=0.005$, it can be clearly identified that the numerical results are coincident very well with the prediction. In this case, the normalized power of the energy source is zero at the phonon band boundaries and it has a maximum value $P/A^2=$1.804 at $\omega=1.913$. This can be understood by the fact that the power at large times is proportional to the phonon group velocity, according to Eq.~(\ref{E appr}), which is zero at the edges of the phonon band and is maximal in the middle of the band.


For the higher driving amplitude $A=0.6$, the nonlinearity comes into play the role. So, the results deviate from the prediction of the linear theory. Specifically, for the model with hard-type anharmonicity, $p/A^2$ appears to increase near the upper edge of the phonon band and to decrease near the lower edge, while in the case of soft-type anharmonicity an opposite tendency is observed.

\begin{figure}
\includegraphics[width=9.0cm]{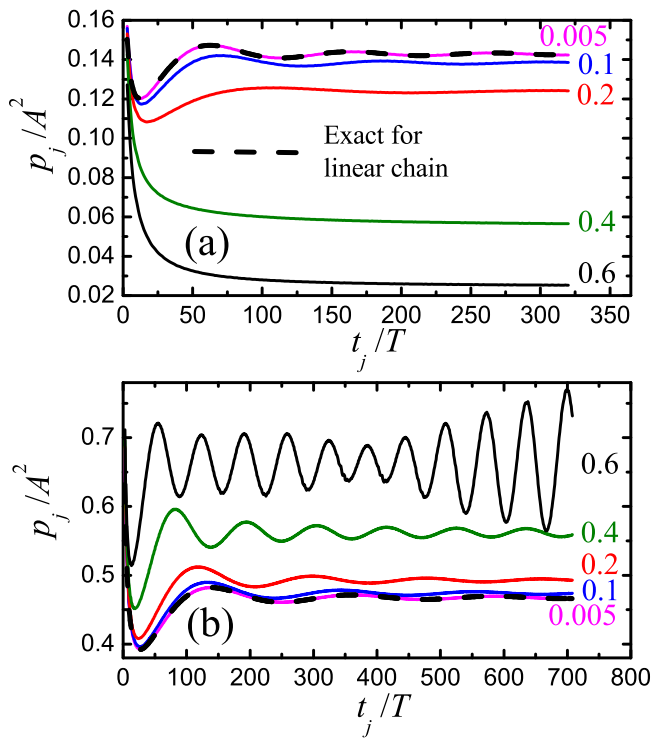}
\caption{Hard-type anharmonicity: Power of the energy source $p_j$ defined by Eq.~(\ref{PowerT}) and normalized by $A^2$ as a function of the driving period number, $j=t_j/T$, for different driving amplitudes
$A$, as indicated for each curve. Driving frequency is inside the phonon band and it is close to (a) the lower edge, $\omega=\omega_{\min}+0.01$, and (b) the upper edge, $\omega=\omega_{\max}-0.01$.} \label{Pjhard}
\end{figure}

\begin{figure}
\includegraphics[width=9.0cm]{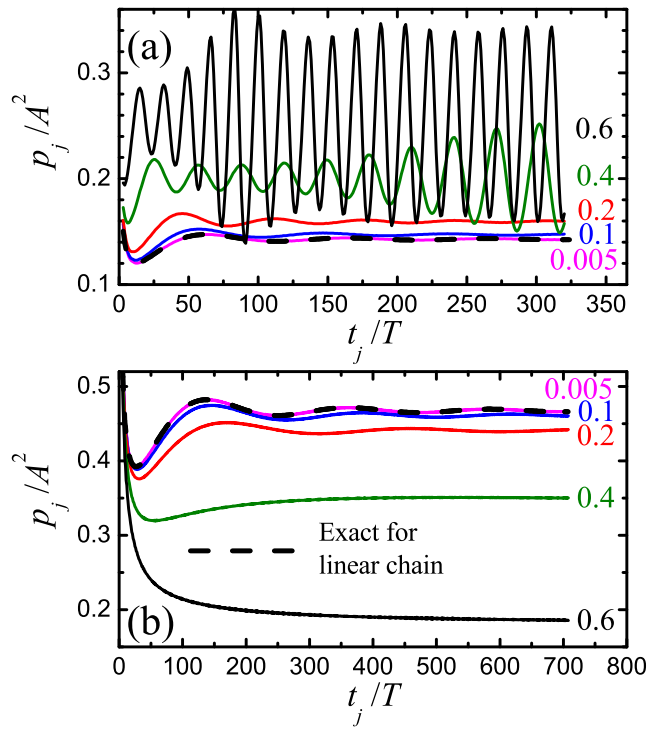}
\caption{The same as that shown in Fig.~\ref{Pjhard}, but for the case of soft-type anharmonicity.} \label{Pjsoft}
\end{figure}

In order to explain this observation, we focus on the driving frequencies within the phonon band but close to its edges. To get a deeper insight of the effect of anharmonicity on energy transfer to the chain from the energy source, we plot $p_j$ defined by Eq.~(\ref{PowerT}) as a function of $j$ for a series of driving amplitudes and driving frequencies $\omega=\omega_{\min}+0.01$ and $\omega=\omega_{\max}-0.01$. Results for the hard-type and soft-type anharmonicity are presented in Figs.~\ref{Pjhard} and \ref{Pjsoft}, respectively, where the panel (a) gives the result of driving frequency close to the lower edge of the phonon band, while panel (b) provides that close to the upper edge of the band. The prediction from the linear chain is plotted with the thick dashed line for a comparison. As can be seen, for $A=0.005$, both the prediction and the simulation results are overlapped as well.


Detailed comparison of Fig.~\ref{Pjhard} and Fig.~\ref{Pjsoft}, one can easily find that both Fig.~\ref{Pjhard}(a) and Fig.~\ref{Pjsoft}(b) show the decrease of the normalized source power with increasing driving amplitude, while a qualitatively different picture is seen in Fig.~\ref{Pjhard}(b) and Fig.~\ref{Pjsoft}(a). The former cases are those that in the lower (upper) edge of the phonon spectrum for hard- (soft-) type anharmonicity, in which after a transient period, $p_j/A^2$ decreases down to a constant value with the increase of $A$. The latter cases then correspond to that, in the upper (lower) edge of the phonon spectrum for hard- (soft-) type anharmonicity, where generally, $p_j/A^2$ shows not a decrease but an increase with $A$, and more importantly, at large $A$ and a long time, $p_j/A^2$ does not approach a constant value but shows quasi-periodic oscillation behaviors.

The formal observations [Fig.~\ref{Pjhard}(a) and Fig.~\ref{Pjsoft}(b)] are understandable and trival, since the driving frequency is close to the edge of the phonon band, while the latter results are interesting, and suggest new underlying mechanisms. Here, we argue that the increase of the power together with quasi-periodic oscillations shown in the Fig.~\ref{Pjhard}(b) and Fig.~\ref{Pjsoft}(a) at large driving amplitudes is related to excitation of moving DBs, since the driving frequency is close to the DB's frequency, while in the cases of Fig.~\ref{Pjhard}(a) and Fig.~\ref{Pjsoft}(b), the driving frequency is far from DB's frequency and DBs are not excited, so a qualitatively different picture can be seen. We will present analytic demonstration for the existence of mobile DBs in Sec.~\ref{Sec:IV}. Before doing that we first provide more details about the transfer of the energy in the chain to further support the arguments.



%
\begin{figure}
\includegraphics[width=9.0cm]{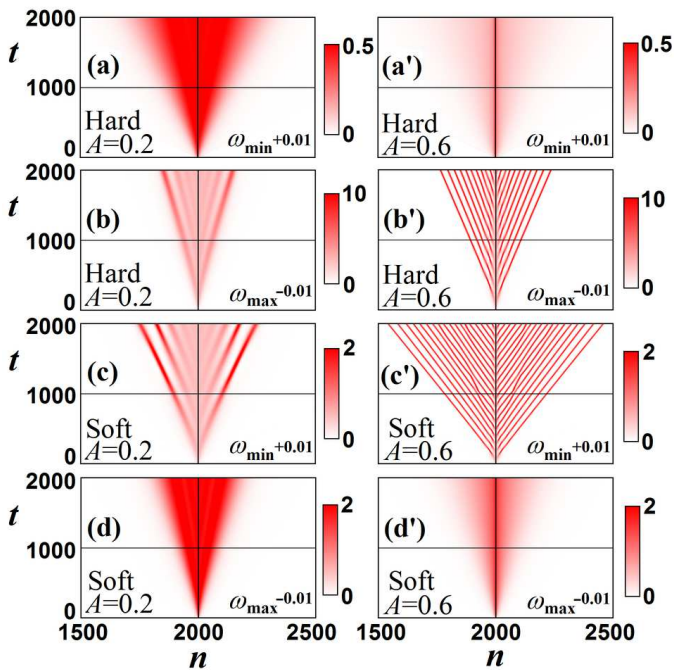}
\caption{Contour plots showing the space-time evolution of normalized energy of the particles, $e_n/A^2$. Inteser colors correspond to higher energies, according to the color bar shown at right of each panel. Particle located at $n=2000$ is driven. (a,a',b,b') Hard-type anharmonicity. (c,c',d,d') Soft-type anharmonicity. (a,a',c,c') Driving frequency is close to the lower edge of the phonon band. (b,b',d,d') Driving frequency is close to the upper edge of the phonon band. Left (right) panels correspond to the driving amplitude $A=0.2$ ($A=0.6$).}
\label{CountorPlot}
\end{figure}
\begin{figure}
\includegraphics[width=9.0cm]{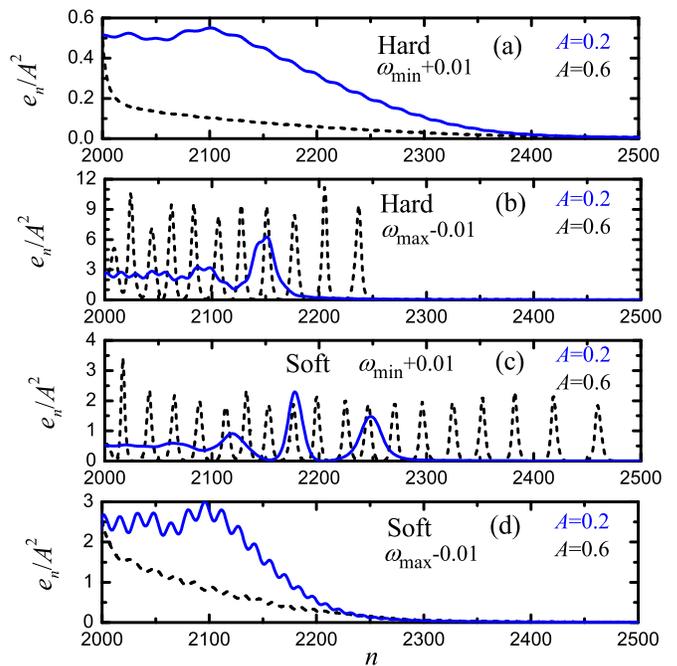}
\caption{Normalized energy of the particles at the end of the numerical run at $t=t_{\max}=2000$. Particle $n=2000$ is driven. A half of the picture is shown due to the mirror symmetry with respect to the driven site. (a,b) Hard-type anharmonicity. (c,d) Soft-type anharmonicity. (a,c) Driving frequency is close to the lower edge of the phonon band. (b,d) Driving frequency is close to the upper edge of the phonon band. Blue solid lines show the results of relatively small driving amplitude $A=0.2$, while black dashed lines correspond to the result of $A=0.6$, when nonlinearity comes into playing the role. Trains of DBs moving away from the energy source can be seen in (b,c), when the driving frequency is close to DB frequency and the driving amplitude is sufficiently large.}
\label{EnerDistr}
\end{figure}

The energy of per particle is usually defined by
\begin{equation}\label{en}
e_n=\frac{m\dot{u}_i^2}{2}+\frac{1}{2}V(u_{n}-u_{n-1})+\frac{1}{2}V(u_{n+1}-u_{n})+U(u_{n}).
\end{equation}
In Fig.~\ref{CountorPlot} a contour plot of the normalized (normalized by $A^2$) total energy of all particles (at $n=2000$, the location of the source), i.e., $e_n/A^2$, during the simulation run up to $t=2000$ is presented. Here, intenser colors are used to correspond to the results of higher energy, according to the color bar. Results of the hard-type anharmonicity are presented in panels (a,a',b,b'), and the counterparts of soft-type anharmonicity are in (c,c',d,d'). In (a,a',c,c') the driving frequency is close to the lower edge of the phonon band, while in (b,b',d,d') it is close to the upper edge of the phonon band. Left (right) panels correspond to the driving amplitude $A=0.2$ ($A=0.6$). 

Indeed, as expected, Fig.~\ref{CountorPlot} shows that the energy flow of panels (a,a',d,d') are qualitatively different from those in panels (b,b',c,c'). This is basically consistent with our above conjecture. In panels (a,a',d,d'), the normalized energy is better accepted by the chain at {\em smaller} driving amplitudes, while in panels (b,b',c,c') the opposite result is true. In particular, in panels (a,a',d,d') the energy radiated by the source is distributed smoothly, while in panels (b,b',c,c') the energy distribution is highly non-uniform, which is better seen for larger driving amplitude in panels (b',c').

We thus present the distribution of $e_n/A^2$ in a more quantitative way. Toward this aim, we show $e_n/A^2$ simulation run of $t=t_{\max}=2000$ and plot it in Fig.~\ref{EnerDistr} with panels (a,b) hard-type anharmonicity, panels (c,d) for soft-type anharmonicity. Remind again, the driven particle is located in the middle of the chain ($n=2000$), and thus only a half of the picture is shown since the energy from the source is emitted symmetrically in both directions, as was evident in Fig.~\ref{CountorPlot}. The same as above, in panels (a,c) the driving frequency is $\omega=\omega_{\min}+0.01$, while in panels (b,d) it is $\omega=\omega_{\max}-0.01$. In each case the results are compared for the relatively small driving amplitude $A=0.2$ (blue solid line) and sufficiently large driving amplitude $A=0.6$ (black dashed line), where the effect of nonlinearity becomes noticeable (see also Figs.~\ref{Pjhard} and \ref{Pjsoft}). 

For the driving frequency far from the DB's frequency [see Fig.~\ref{EnerDistr} (a) and (d)] $e_n/A^2$ is larger for smaller $A$, in line with the results shown in Fig.~\ref{Pjhard}(a) and Fig.~\ref{Pjsoft}(b). The opposite fact is true for the driving frequency close to the DB's frequency [see Fig.~\ref{EnerDistr} (b) and (c)], as it was already concluded in Fig.~\ref{Pjhard}(b) and Fig.~\ref{Pjsoft}(a).

More-interestingly, in the case of the driving frequency close to the DB's frequency, we almost recover the quasi-oscillation behavior as shown in Fig.~\ref{Pjhard}(b) and Fig.~\ref{Pjsoft}(a). In this case, the energy distribution is in the form of a series of peaks corresponding very similarly to a train of moving DBs emitted by the energy source. Regarding the difference between the results of Figs.~\ref{EnerDistr} (b) and (c), we point out that this may because, DBs emitted at driving amplitude $A=0.6$ propagate faster than the small-amplitude waves emitted at $A=0.2$.

\begin{figure}
\includegraphics[width=9.0cm]{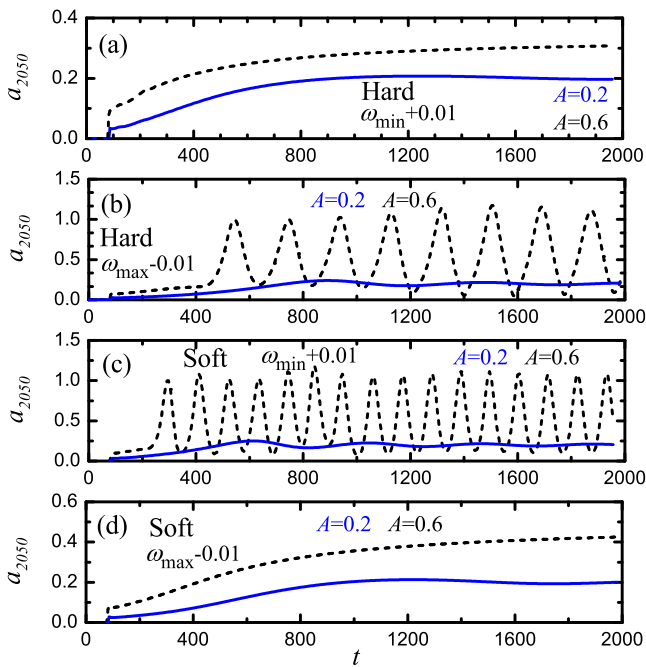}
\caption{Vibration amplitude of $n=2050$ particle as a function of time (driven particle is $n=2000$). (a,b) Hard-type anharmonicity. (c,d) Soft-type anharmonicity. (a,c) Driving frequency is close to the lower edge of the phonon band. (b,d) Driving frequency is close to the upper edge of the phonon band. Blue solid (black dashed) lines show the results for $A=0.2$ ($A=0.6$).}
\label{Particle}
\end{figure}

To gain a clearer evidence, we finally analyze the vibration amplitude of one particle located at $n=2050$, which is 50 cites away from the energy source. The vibration amplitude, $a_{2050}$, as a function of time is presented in Fig.~\ref{Particle}, where the relevant parameters are the same as those in Fig.~\ref{EnerDistr}. From Fig.~\ref{Particle}, for small driving amplitude, e.g., $A=0.2$, the results are similar regardless which types of anharmonicity and what values of the driving frequency. In all of the four cases, as time grows, $a_{2050}$ approaches the driving amplitude. This can be understood by the picture that, at small driving amplitudes, the source emits phonons with the amplitude equal to the driving amplitude. Whereas the results for the larger driving amplitude, e.g., $A=0.6$, are sensitive to the parameters. When the driving frequency is far from DB's frequencies [Fig.~\ref{Particle} (a) and (d)], $a_{2050}$ increases with time monotonically, while for the driving frequency close to the DB's frequencies, it is oscillated in time since energies are carried by DBs passing quasi-periodically through this particle. DB's amplitudes in both cases are slightly above $1$, which are obvious larger than the driving amplitude.

Based on Fig.~\ref{Particle}(b) and Fig.~\ref{Particle}(c), further calculations of the instant vibration frequency of the focused particle give that the frequency $\omega=2.243$ for the case of hard-type anharmonicity, which is above the phonon spectrum; while $\omega=0.972$ for the soft-type nonlinearity, which is clearly below the phonon spectrum. Such frequencies regimes are just within the DB's frequencies outside the linear phonon band in each model, further supporting the fact that they just correspond to DBs.

\section {Discrete breathers} \label{Sec:IV}

We now provide information on DBs. We will first focus on the properties of standing DBs and then consider their mobility.

\subsection {Standing discrete breathers} \label{Sec:IVA}

To excite a standing DB, the following ansatz was used for the case of hard-type nonlinearity
\begin{equation}\label{DBh}
u_n(0)=\frac{(-1)^{n}A_{\rm DB}}{\cosh[{\theta (n-N/2)}]}, \quad \dot{u}_n(0)=0,
\end{equation}
while for the soft-type anharmoncity, we adopt
\begin{equation}\label{DBs}
u_n(0)=\frac{A_{\rm DB}}{\cosh[{\theta (n-N/2)}]}, \quad
\dot{u}_n(0)=0.
\end{equation}
Here $A_{\rm DB}$ and $\theta$ are the DB's amplitude and inverse width, respectively. DB is centered on the middle particle of the chain, $n=N/2$. We stress that Eqs.~(\ref{DBh}) and (\ref{DBs}) are not the exact solutions to Eq.~(\ref{EMo}), but they produce fairly good initial conditions for DBs. For the chosen $A_{\rm DB}$, we find $\theta$ by using the trial and error method~\cite{Kistanov} minimizing the oscillations of the DB amplitude in simulations. After $\theta$ is determined, we then calculate DB frequency, $\omega_{\rm DB}$, and its total (kinetic plus potential) energy, $E_{\rm DB}$. These results are presented in Table~\ref{T1} for a set of DB amplitudes for both hard- and soft-type nonlinearities.

\begin{table}[!hbp]
\begin{centering}
\caption{\label{T1} Parameters of standing DB.}
\begin{tabular}{ p{1.5cm}||p{1.5cm}|p{1.5cm}|p{1.5cm}}
 \multicolumn{4}{c}{Hard-type anharmonicity} \\
 \hline
 $\,A_{\rm DB}$ & $\,\,\,\theta$ & $\,\omega_{\rm DB}$ & $\,E_{\rm DB}$ \\
 \hline
 $\,$0.5   & $\,$0.126  & $\,$2.240 & $\,$  9.954  \\
 $\,$0.75  & $\,$0.190  & $\,$2.244 & $\,$  14.84  \\
 $\,$1.0   & $\,$0.257  & $\,$2.250 & $\,$  19.59  \\
 $\,$1.25  & $\,$0.326  & $\,$2.259 & $\,$  24.27  \\
 $\,$1.5   & $\,$0.398  & $\,$2.270 & $\,$  28.72  \\
 $\,$1.75  & $\,$0.475  & $\,$2.283 & $\,$  32.95  \\
 $\,$2.0   & $\,$0.560  & $\,$2.299 & $\,$  36.75  \\
 $\,$2.25  & $\,$0.658  & $\,$2.318 & $\,$  39.87  \\
 $\,$2.5   & $\,$0.783  & $\,$2.339 & $\,$  41.64  \\
 $\,$2.75  & $\,$0.933  & $\,$2.369 & $\,$  42.59  \\
 $\,$3.0   & $\,$1.097  & $\,$2.403 & $\,$  43.63  \\
 \hline
 \multicolumn{4}{c}{$\,$} \\
 \multicolumn{4}{c}{Soft-type anharmonicity} \\
 \hline
 $\,A_{\rm DB}$ & $\,\,\,\theta$ & $\,\omega_{\rm DB}$ & $\,E_{\rm DB}$ \\
 \hline
 $\,$0.5   & $\,$0.125  & $\,$0.992 & $\,$  1.991  \\
 $\,$0.75  & $\,$0.186  & $\,$0.983 & $\,$  2.969  \\
 $\,$1.0   & $\,$0.246  & $\,$0.969 & $\,$  3.928  \\
 $\,$1.25  & $\,$0.305  & $\,$0.952 & $\,$  4.861  \\
 $\,$1.5   & $\,$0.361  & $\,$0.932 & $\,$  5.770  \\
 $\,$1.75  & $\,$0.414  & $\,$0.908 & $\,$  6.656  \\
 $\,$2.0   & $\,$0.464  & $\,$0.882 & $\,$  7.513  \\
 $\,$2.25  & $\,$0.511  & $\,$0.854 & $\,$  8.334  \\
 $\,$2.5   & $\,$0.551  & $\,$0.824 & $\,$  9.170  \\
 $\,$2.75  & $\,$0.585  & $\,$0.793 & $\,$  10.01  \\
 $\,$3.0   & $\,$0.613  & $\,$0.764 & $\,$  10.87  \\
 \hline
\end{tabular}
\end{centering}
\end{table}

Table~\ref{T1} tells us that, with the increase of DB's amplitude, the degree of its spatial localization, characterized by $\theta$, increases. The same is true for DB's energy. While DB's frequency increases (decreases) with amplitude being above (below) the phonon band for hard-type (soft-type) anharmonicity. 

\begin{figure}
\includegraphics[width=9.5cm]{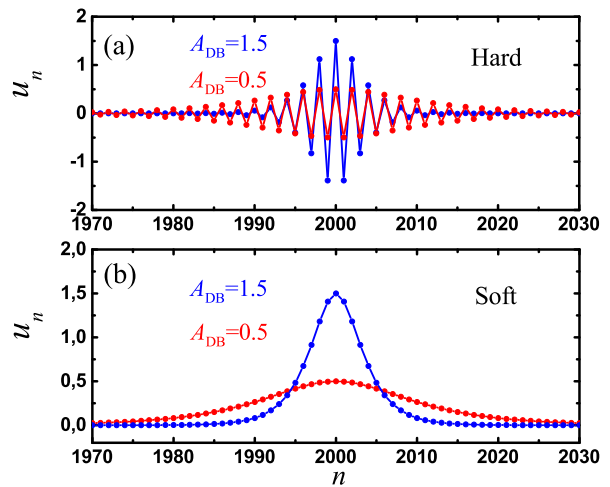}
\caption{Standing DBs profiles for two different amplitudes $A_{\rm DB}=0.5$ and $A_{\rm DB}=1.5$ for the case of (a) hard-type and (b) soft-type nonlinearities.} \label{fig44}
\end{figure}

Based on Table~\ref{T1}, Fig.~\ref{fig44} further plots several typical DBs profiles for hard-type [see Fig.~\ref{fig44}(a)] and soft-type [see Fig.~\ref{fig44}(b)] anharmonicities. Here, we only plot the DBs at the oscillation phase when particles have largest displacements. A comparison of the results of $A_{\rm DB}=0.5$ and $ A_{\rm DB}=1.5$ indicates that, DBs with larger amplitude are more localized. In addition to this common feature, the DBs profiles for both types of anharmonicities are different, i.e., in the case of hard-type (soft-type) nonlinearity DB has a staggered (smooth) shape.

\subsection {Moving discrete breathers} \label{Sec:IVB}

Moving DBs were excited by using the following physically motivated ansatz~\cite{Kistanov}. For hard-type nonlinearity, it has the form
\begin{equation} \label{Qmh}
u_n(t)=\frac{(-1)^n A_{\rm DB}\cos[\omega_{\rm DB} t+\delta(n-x_0)]}{\cosh[\theta(n-x_0)]},
\end{equation}
while for soft-type nonlinearity it reads
\begin{equation} \label{Qms}
u_n(t)=\frac{A_{\rm DB}\cos[\omega_{\rm DB} t+\delta(n-x_0)]}{\cosh[\theta(n-x_0)]}.
\end{equation}
Here, $\delta$ is a free parameter which characterizes DB's velocity, $v_{\rm DB}$, in case when it is mobile. For example, for $\delta=0$ DB's velocity is zero, and both Eqs.~\eqref{Qmh} and \eqref{Qms} essentially reduce to the origin ones, Eqs.~\eqref{DBh} and \eqref{DBs}, respectively. Change of the sign of $\delta$ results in the change of the sign of DB's velocity.

It should be pointed out that Eqs.~(\ref{Qmh}) and (\ref{Qms}) do not describe exact moving DBs, but they give very good approximate solutions for moving DBs in the case of not very high DB amplitude.
\begin{figure}
\includegraphics[width=8.0cm]{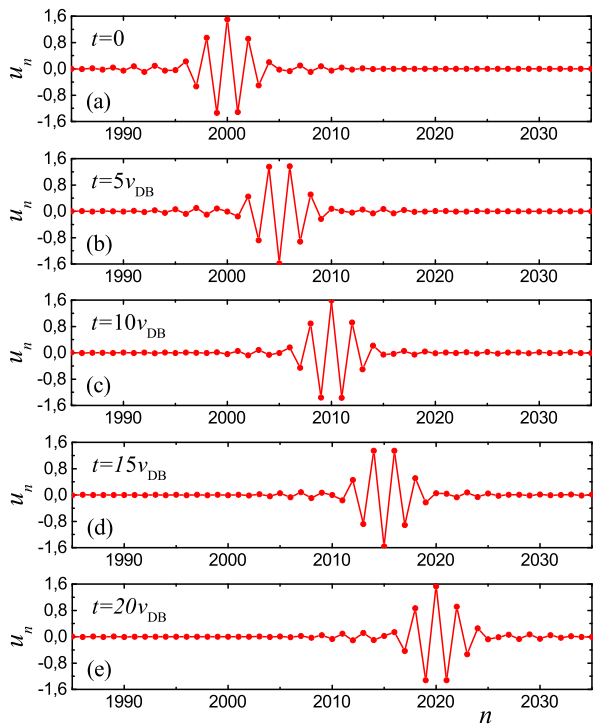}
\caption{Moving DB's profile in the case of hard-type anharmonicity. Parameters of the ansatz Eq.~(\ref{Qmh}) are $A_{\rm DB}$=1.5, $\theta$=0.398, $\omega_{\rm DB}$=2.270, $x_0=2000$, $\delta=0.3$. DB's velocity $v_{\rm DB}$=0.1303. Time is indicated in each panel.} \label{MovingDBhard}
\end{figure}
\begin{figure}
\includegraphics[width=8.0cm]{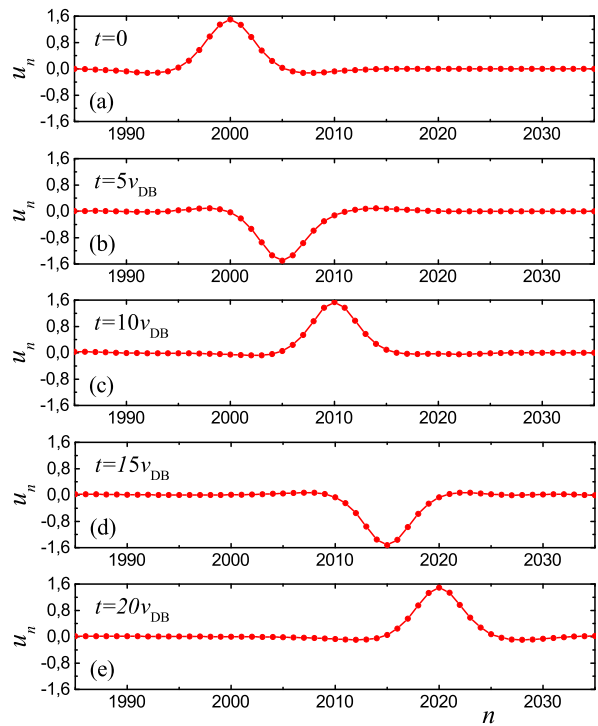}
\caption{Moving DB's profile in the case of soft-type anharmonicity. Parameters of the ansatz Eq.~(\ref{Qms}) are $A_{\rm DB}$=1.5, $\theta$=0.361, $\omega_{\rm DB}$=0.932, $x_0=2000$, $\delta=0.3$. DB's velocity $v_{\rm DB}$=0.2838. Time is indicated in each panel.} \label{MovingDBsoft}
\end{figure}
\begin{figure}
\includegraphics[width=9.5cm]{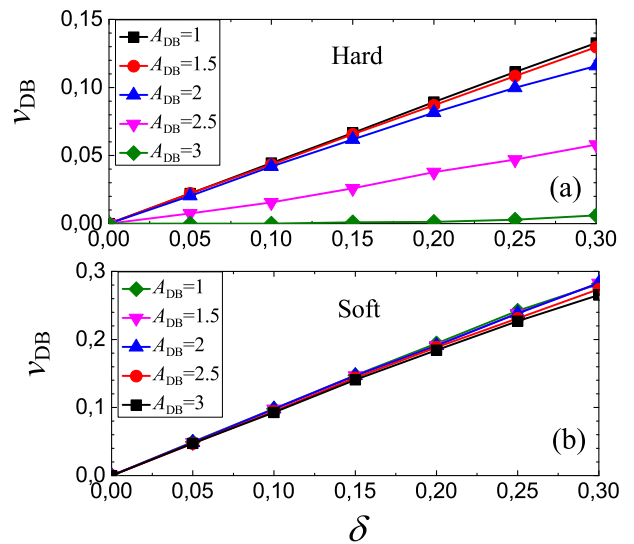}
\caption{Velocity $v_{\rm DB}$ of DBs as a function of $\delta$ for (a) hard-type and (b) soft-type nonlinearities. DB's amplitude is indicated in the legends. Other DB's parameters are taken from Table~\ref{T1}.} \label{vDB}
\end{figure}

We use Eqs.~\eqref{Qmh} and \eqref{Qms} with different values of $\delta$ for setting the initial conditions taking other DB parameters from Table~\ref{T1}. Two typical examples of moving DBs evolution are shown in Fig.~\ref{MovingDBhard} and Fig.~\ref{MovingDBsoft} for the hard-type and soft-type nonlinearities, respectively. Here, we choose $\delta=0.3$, $A_{\rm DB}=1.5$, $x_0=2000$. Other parameters are taken from Table~\ref{T1}. The measured DB's velocity is $v_{\rm DB}$=0.1303 in Fig.~\ref{MovingDBhard} and $v_{\rm DB}$=0.2838 in Fig.~\ref{MovingDBsoft}. For these chosen parameters, DBs propagate at constant velocities practically radiating no energy.

Velocity of DB is measured and presented as a function of $\delta$ in Fig.~\ref{vDB}(a) and (b) for the chains with hard-type and soft-type anharmonicity, respectively.
Different lines show the results for different DB amplitudes, $A_{\rm DB}$, as indicated in the legends. 

In fact, in our calculations, we have examined DB's velocities for different $A$ in detail. The relevant results are presented in Fig.~\ref{vDB}. It can be seen that, for hard-type anharmonicity [see Fig.~\ref{vDB}(a)], DBs with relatively small amplitudes ($A_{\rm DB}<2$) have velocities nearly proportional to $\delta$ within the range of $ |\delta| \le 0.3$ considered here. Such DBs are highly mobile. We have checked that they move through entire computational cell of 4000 particles with nearly constant velocity and practically radiating no energy. However, for the cases of $A_{\rm DB}=2.5$, the increase of DB's velocity with $\delta$ is slower than that for smaller amplitudes. While propagating, it radiates small-amplitude waves and its velocity gradually decreases. For this reason, we measured the DB's velocity at $t=300$. Finally, DBs with even higher amplitudes ($A_{\rm DB}\ge 3$) are trapped by the lattice and no longer move for any value of $\delta$. On the other hand, as evidenced by Fig.~\ref{vDB}(b), DBs in the lattice with soft-type nonlinearity are highly mobile for all the considered amplitudes up to $A_{\rm DB}=3$. 

With above information, now we try to relate the energy transfer process due to the ac driving to properties of moving DBs. As it was already mentioned in Sec.~\ref{Sec:IIIB} that, DBs emitted by the driven particle at driving amplitude $A=0.6$ have amplitudes slightly above $A=1$ for both hard- and soft-type nonlinearities [see dashed lines in Fig.~\ref{Particle}(b) and (c), respectively]. Velocity of DBs emitted by the driven particle at driving amplitude $A=0.6$ can be estimated from Fig.~\ref{CountorPlot}(b') and (c') for hard- and soft-type nonlinearities, respectively, and they are found to be about $0.12$ and $0.23$, respectively. These relevant parameters (amplitude and velocity) naturally share similar values typical for moving DBs considered here, thus supporting our conjecture that, indeed, moving DBs are responsible for the mechanism.

\section{Conclusions}
\label{Sec:V}

Two typical chains of harmonically coupled particles placed in the six-order polynomial on-site potentials of hard-type and soft-type nonlinearities have been analyzed. Firstly, energy transfer to the chain from one harmonically driven particle was analyzed for different driving amplitudes and for two driving frequencies within the phonon band and close to the upper and lower edges of the band. Secondly, properties of discrete breathers (DBs) were studied.

Our main findings are summarized as follows:

\begin{itemize}
\item An exact solution for the power of energy source in the form of one particle moving according to the harmonic law with amplitude $A$ and frequency $\omega$ in harmonic chain has been obtained, see Eq.~(\ref{sol w_n}). From this solution, the power of energy source increases proportionally to $A^2$. For large times, the power of energy source normalized to $A^2$ is proportional to $\omega^2$ and proportional to the group velocity of phonons with frequency $\omega$. This means that the power at large times vanishes for driving frequencies at the edges of the phonon band, where phonon group velocity vanishes.

\item For the considered nonlinear models, driving with the amplitude $A\le 0.2$ can be described by the linear theory quite well. For driving amplitudes $A>0.4$, the effect of nonlinearity should be taken into account. 

\item When driving frequency is far from the DB's frequency and close to the edge of the phonon spectrum, increase in the driving amplitude results in the reduction of the power, and as time increases, power of the energy source approaches a constant value, see Fig.~\ref{Pjhard}(a) and Fig.~\ref{Pjsoft}(b).

\item When the driving frequency is close to the DB's frequency and also close to the edge of the phonon spectrum, increase in the driving amplitude results in the increase of the power, and the power oscillates with time quasi-periodically, see Fig.~\ref{Pjhard}(b) and Fig.~\ref{Pjsoft}(a). These oscillations reflect the emission of DBs moving away from the energy source, see Fig.~\ref{EnerDistr} (b) and (c). In the previous works \cite{GL2002,Rings2004,Evazzade,Aranson,Caputo,Khomeriki}, emission of DBs by driving with frequencies {\em outside} the phonon band was reported, but here we demonstrate that they can also be excited with driving frequencies {\em inside} the phonon band close to the DB frequency.

\end{itemize}

Overall, we have demonstrated that in the case of moderate driving amplitudes and driving frequencies close to the edges of phonon band (within the band), DBs enhance energy transfer to the chain from the harmonically driven particle. This result contributes to our understanding of dynamics of nonlinear chains under external driving and uncovers the role of DBs in such systems.
-

\section*{Acknowledgments}

Stay of D.S. at IMSP RAS was partly supported by the Russian Science Foundation, grant No. 14-13-00982. Work of D.X. was supported by the National Natural Science Foundation of China (Grant No. 11575046), the Natural Science Foundation of Fujian Province, China (Grant No. 2017J06002), and the Qishan Scholar Research Fund of Fuzhou University, China. Work of V.A. Kuzkin was supported by the Russian Science Foundation (RSCF grant No. 17-71-10213). Work of A.M. Krivtsov was supported by the Russian Foundation for Basic Research (RFBR grant No. 16-29-15121). S.V.D. was supported by the Russian Science Foundation, grant No. 16-12-10175.

\end{document}